\documentstyle[12pt]{article}
\input mssymb
\font\teneuf=eufm10
\font\seveneuf=eufm7
\font\fiveeuf=eufm5
\newfam\frakturfam
\textfont\frakturfam\teneuf
\scriptfont\frakturfam\seveneuf
\scriptscriptfont\frakturfam\fiveeuf
\def\frak{\fam\frakturfam\teneuf}

\newcommand{\QED}{\hspace{0.2in}\vrule width 6pt height 6pt depth 0pt
\vspace{0.1in}}

\newcommand{\Proof}{{\em Proof} \hspace{0.2in}}
\newcommand{\tag}[1]{\eqno(#1)}

\newtheorem{theorem}{Theorem}[section]

\newtheorem{proposition}[theorem]{Proposition}

\title{Quantum symmetric spaces}
\author{J.Donin and S.Shnider\\
Department of Mathematics, Bar-Ilan University}
\date{}
\begin{document}
\maketitle
\begin{abstract}
Let $G$ be a semisimple Lie group, ${\frak g}$ its Lie algebra.
For any symmetric space $M$ over $G$ we construct a new (deformed)
multiplication in the space $A$ of smooth functions on $M$. This
multiplication is invariant under the action of the Drinfeld--Jimbo
quantum group $U_h{\frak g}$ and is commutative with respect to an
involutive operator $\tilde{S}: A\otimes A \to A\otimes A$.
Such a multiplication is unique.

Let $M$ be a k\"{a}hlerian symmetric space with the canonical Poisson
structure. Then we construct a $U_h{\frak g}$-invariant multiplication
in $A$ which depends on two parameters and is a quantization of
that structure.
\end{abstract}

\def\bC{{\Bbb C}}
\def\bR{{\Bbb R}}
\def\g{{\frak g}}
\def\w{\wedge}
\def\al{\alpha}
\def\t{\otimes}
\def\ff{\varphi}
\def\si{\sigma}
\def\c{{\frak c}}
\def\k{{\bf k}}
\def\bh{[[h]]}
\def\na{\nabla}
\def\De{\Delta}
\def\VM{\mbox{Vect}(M)}
\def\CM{C^\infty (M)}
\def\br#1#2{\{#1,#2\}}
\def\R{{\rm Rep}}
\def\bd{\br{\cdot}{\cdot}}

\section{Introduction}
\label{s0}
Let $G$ be a semisimple Lie group, $\g$ its Lie algebra,
and $r\in\wedge^{\t 2}\g$ the Drinfeld--Jimbo classical $R$-matrix
(see Section \ref{s1}).
Suppose $H$ is a closed subgroup of $G$ and $M=G/H$.
Then the action of $G$ on $M$ defines a mapping $\rho:\g\to\VM$.
So, the element $(\rho\t\rho)(r)$ induces a bivector field on $M$
which determines a bracket
(biderivation)  $\bd$ on the algebra $\CM$ of smooth functions on $M$.
In some cases this will be satisfy the Jacobi identity and thus
define a Poisson bracket which we will call
an $R$-matrix Poisson bracket. It is easy to see that the bracket
may be degenerate at some points of $M$.
The natural question arises whether that bracket can be quantized.

The first case when $\bd$ is a Poisson bracket is when the Lie algebra
of $H$ contains a maximal nilpotent subalgebra.
In \cite{DGM} it is proven that in this case there exists a
quantization of $\bd$, i.e., there is an associative multiplication
$\mu_h$ in $\CM$ of the form
$$\mu_h=m+h\bd+\sum_{i=2}^\infty h^i\mu_i(\cdot,\cdot)=m+h\bd+o(h),$$
 where $m$ is the
usual multiplication in $\CM$ and $\mu_i(\cdot,\cdot)$ are
bidifferential operators.
Moreover, this multiplication will be invariant under action
of the Drinfeld--Jimbo quantum group $U_h\g$. This means that
$\mu_h$ satisfies the condition
$$x\mu_h(a,b)=\mu_h\tilde{\De}_h(x)(a\t b),$$
where $a,b\in\CM$, $x\in U_h\g$, and $\tilde{\De}_h$ is the
comultiplication in $U_h\g$ (here we use the presentation of $U_h\g$
with multiplication as in $U\g[[h]]$, see Section 3).
In \cite{DG1} it is shown that in such a
way one can obtain the $U_h\g$-invariant quantization of the algebra
of holomorphic sections of line bundles over the flag manifold of $G$.

In the present paper we consider the case when $M$ is a symmetric
space.
Our first result is that in this case $\bd$ will also be a Poisson
bracket and there is a $U_h\g$-invariant quantization of this bracket.
Moreover, such a quantization is unique up to isomorphism.

Suppose now that $M$ is equipped with a $G$-invariant Poisson bracket
$\bd_{inv}$.
Our second result is that in this case there exists a simultaneous
$U_h\g$-invariant
quantization, $\mu_{\nu,h}$, of both these brackets
in the form
$$\mu_{\nu,h}=m+\nu\bd_{inv}+h\bd+o(\nu,h),$$
where $o(\nu,h)$ includes all terms of total powers $\ge 2$ in $\nu,h$
with bidifferential operators as coefficients.
This is the case, for example, when $M$ is a k\"{a}hlerian symmetric
space.
Then $\bd_{inv}$ coincides with the Kirillov bracket which is dual to
the K\"{a}hler form on $M$. This bracket is
nondegenerate, and Melotte \cite{Me} has  proved
and one can prove that there exists
a deformation quantization of the Kirillov bracket, $\mu_\nu$,
that is invariant under
$G$ and $U\g$. The existence of such a
quantization can be also proven using the methods of the present
paper.
Thus, one may consider the multiplication
$\mu_{\nu,h}$ as such a quantization of the Kirillov bracket which
is invariant under the action of the quantum group $U_h\g$.

Note that the Kirillov bracket is also generated by $r$ in the following
way. Let $\bd'$ be a bracket on $C^\infty(G)$ generated by
the left-invariant extension of $r$ as a bivector field on $G$.
Using the projection $G\to G/H=M$ we can consider $\CM$ as
a subalgebra of $C^\infty(G)$. One can  check that $\CM$ is
invariant under $\bd'$ if $H$ is a Levi subgroup. For such $H$
the difference $\bd-\bd'$ gives a Poisson bracket on $M$, the so-called
Sklyanin--Drinfeld Poisson bracket. The quantization of this Poisson
bracket is given in \cite{DG2}. In case $M$ is a symmetric space
the bracket $\bd'$ will be a Poisson one itself and coincides
with the Kirillov bracket $\bd_{inv}$ (see \cite{DG2}).
In \cite{GP} there is given a classification of all orbits in the coadjoint
representation of $G$ on which $r$ induces the Poisson bracket.

The authors wish to thank D.Gurevich for stimulating discussions
about subject of the paper.

\section{$R$-matrix Poisson brackets on symmetric spaces}
\label{s1}
Let $\g$ be a simple Lie algebra over the field of complex numbers
$\bC$. Fix a Cartan decomposition of $\g$ with corresponding
root system $\Omega$ and choice of positive roots, $\Omega^+$.
We consider the Drinfeld--Jimbo classical $R$-matrix
$$r=\sum_{\al\in\Omega^+} X_\al\w X_{-\al}\in\w^2\g, $$
where $X_\al$ are the elements from the Cartan--Chevalley basis
of $\g$ corresponding to $\Omega$, and $\Omega^+$ denotes the set
of positive roots. We shall use the notation  $r=r_1\t r_2$
 as a shorthand for $\sum_i r_{1i}\t r_{2i}$ in denoting  this $R$-matrix.
The same convention of suppressing the summation sign and the index of
summation will be used throughout the paper.

This $r$ satisfies the so-called modified classical Yang-Baxter
equation which means that the Schouten bracket of $r$ with itself
is equal to an invariant element $\ff\in \w^3 \g$:
$$[r,r]_{Sch}=[r^{12},r^{13}]+[r^{12},r^{23}]+[r^{13},r^{23}]=\ff.\tag{1}$$
Here we use the usual notation: $r^{12}=r_1\t r_2\t 1$,
$r^{13}=r_1\t 1\t r_2$, and so on.
Note that any invariant element in $\w^3\g$ is dual up to a
multiple to the three-form $(x,[y,z])$ on $\g$, where
$(\cdot,\cdot)$ denotes the Killing form. Therefore, $\ff$ will be
also invariant under all automorphisms of the Lie algebra $\g$.

The $R$-matrix $r$ obviously satisfies the following conditions:
a) it is invariant under the Cartan subalgebra $\c$, and b)
$\theta r=-r$ where $\theta$ is the Cartan involution of $\g$,
$\theta X_\al=-X_{-\al},\ \theta|_{\c}=-1$.
 These conditions determine $r$ uniquely
up to a multiple (see \cite{SS} \S 11.4).

In case $\g$ is a semisimple Lie algebra with a Cartan decomposition,
let $r\in \w^2\g$ satisfy the
equation (1) for some invariant $\ff\in \w^3\g$ and the previous
conditions a) and b). Then $r$ will be a linear combination of the
Drinfeld--Jimbo $R$-matrices on the simple components of $\g$.
We will also call such $r$ the Drinfeld--Jimbo $R$-matrix.

Let $\g_{\bR}$ be a real form of a semisimple (complex) Lie algebra $\g$,
and $G$ a connected Lie group with $\g_{\bR}$ as its Lie algebra.
Suppose $\sigma$ is an involutive automorphism of $G$, and $H$ is
a subgroup of $G$ such that $G^\sigma_0\subset H\subset G^\sigma$,
where $G^\sigma$ is the set of fixed points of $\sigma$ and $G^\si_0$
is the identity component of $G^\si$. The automorphism $\si$ induces
an automorphism of the both Lie algebras $\g_{\bR}$ and $\g$ which
we will also denote by $\si$. Thus, the space of left cosets
$M=G/H$ is a symmetric space (see \cite{He}).
We denote by $o$ the image of unity by the natural projection
$G\to M$. The mapping $\tau: M\to M, gH\mapsto \si(g)H$, is well defined
and has $o$ as an isolated fixed point, therefore, the differential
$\dot{\tau}: T_o\to T_o$ of $\tau$ at the point $o$ multiplies
the vectors of the tangent space $T_o$ by $(-1)$.

The action of $G$ on $M$ defines the mapping of $\g_{\bR}$ into
the Lie algebra of real vector fields on $M$,
$\rho:\g_{\bR}\to {\rm Vect}_{\bR}(M)$, that extends to a mapping
$\rho:\g\to \VM$ of the complexification of $\g_{\bR}$ into the Lie
algebra of complex vector fields $\VM$ on $M$.

The mapping $\rho$ induces on $M$ a skew-symmetric bivector field
 in the following way.
The element $\rho(r_1)\t\rho(r_2)\in \w^2 \VM$ (tensor product over
$\bC$  not $\CM$) generates
a bracket on the algebra $\CM$ of smooth complex-valued functions
on $M$, $\{f,g\}=\rho(r_1)f\cdot\rho(r_2)g$, where $f,g\in \CM$
and $\rho(r_1)f$ is the derivative of $f$ along the vector field
$\rho(r_1)$. It is obvious that this defines a skew-symmetric biderivation,
therefore it is defined by a bivector field, i.e., a section of
$\wedge^2$ of the tangent bundle, which we denote by $\rho(r)$.

 From now on we will suppose that the invariant element $\ff\in \w^3\g$
is invariant under $\si$ as well. In case $\g$ is a simple Lie algebra
this will be satisfied automatically.

\begin{proposition}
\label{p1.1}
The bracket $\{\cdot,\cdot\}$ is a Poisson bracket on $M$.
\end{proposition}
\Proof
 Since $\rho(\ff)$ is a $G$-invariant three-vector
field on $M$, therefore it is defined by its value at the point $o$,
$\rho(\ff)_o$. Since $\ff$ is $\si$-invariant, $\rho(\ff)$ has to be
$\tau$-invariant, which implies that
$\dot{\tau}\rho(\ff)_o=\rho(\ff)_o$. But the operator $\dot{\tau}$
acts on $T_o$ by multiplying by $(-1)$, so that
$\dot{\tau}\rho(\ff)_o=-\rho(\ff)_o$. Therefore, $\rho(\ff)=0$.
This means that the Schouten bracket $[\rho(r),\rho(r)]$ is equal to
zero, which is equivalent to the bracket $\{\cdot,\cdot\}$
 satisfying  the Jacobi identity. \QED

We will call the bracket $\br{\cdot}{\cdot}$ an $R$-matrix Poisson
bracket. Note that this bracket is not $\g$-invariant and may be
degenerate in some points of $M$.

Suppose now that there is on $M$ a $\g$-invariant Poisson bracket
$\br{\cdot}{\cdot}_{inv}$. The case will be if the
Poisson structure  on $M$ is dual to
a $G$ invariant symplectic form, as in the case of
a k\"{a}hlerian symmetric space. For example, if $M$ is a hermitian
symmetric space the k\"{a}hlerian form is the imaginary part of the
hermitian form on $M$.

\begin{proposition}
\label{p1.2}
The $R$-matrix and any invariant Poisson brackets are compatible, i.e.
for any $a,b\in\bC$ the bracket
$a\br{\cdot}{\cdot}+b\br{\cdot}{\cdot}_{inv}$ is a Poisson one.
\end{proposition}
\Proof The straightforward computation following from the fact
that $\br{\cdot}{\cdot}$  is expressed in terms of vector fields coming from
$\g$ and $\br{\cdot}{\cdot}_{inv}$ is $G$ invariant. (see \cite{DGM}). \QED

\section{Three monoidal categories}
\label{s2}
\def\cA{{\cal C}}

We recall that a monoidal category is a triple $(\cA,\t,\phi)$
where $\cA$ is a category equipped with a functor $\t:\cA\times\cA\to
\cA$, called a tensor product functor, and a functorial isomorphism
$\phi_{X,Y,Z}:(X\t Y)\t Z)\to X\t (Y\t Z)$
called associativity constraint, which satisfies the pentagon
identity (omitting subscripts on $\phi$), i.e. the diagram
$$
\begin{array}{ccccc}
((X\t Y)\t Z)\t U &\stackrel{{\textstyle \phi}}{\longrightarrow}
& (X\t Y)\t(Z\t U) &\stackrel{{\textstyle \phi}}{\longrightarrow}
& X\t (Y\t (Z\t U))  \\
\phi\t id \downarrow & & & & \uparrow id\t\phi \\

(X\t (Y\t Z))\t U& &
\frac{\quad\quad\quad\quad{\textstyle \phi}\quad\quad\quad\quad}
{\quad\quad\quad\quad\quad\quad\quad\quad\quad}
\!\!\!\longrightarrow& &                     X\t ((Y\t Z)\t U)
\end{array}
\tag{1}
$$
is commutative.

\def\wt{\widetilde}

If $(\wt{\cA},\wt{\t},\wt{\phi})$ is another monoidal category,
then a morphism from $\cA$ to $\wt{\cA}$ is given by a pair
$(\alpha, \beta)$ where $\alpha:\cA\to \wt{\cA}$ is a functor
and $\beta:\alpha(X\t Y)\to \alpha(X)\wt{\t}\alpha(Y)$ is a functorial
isomorphism such that the diagram

$$
\begin{array}{ccccc}
\al ((X\t Y)\t Z) & \stackrel{{\textstyle \beta}}{\longrightarrow}
&\al (X\t\ Y)\wt{\t}\al (Z)
&\stackrel{{\textstyle \beta\wt{\t}id}}{\longrightarrow}
&(\al (X)\wt{\t}\al (Y))\wt{\t}\al (Z)\\
\al(\phi)\downarrow & & & & \downarrow\wt{\phi}\\
\al (X\t (Y\t Z)) & \stackrel{{\textstyle \beta}}{\longrightarrow}
&\al (X)\wt{\t}\al (Y\t Z)
&\stackrel{{\textstyle id\wt{\t}\beta}}{\longrightarrow}
&\al (X)\wt{\t}(\al (Y)\wt{\t}\al (Z))
\end{array}
\tag{2}
$$
is commutative.

The morphism $(\al,\beta)$ of monoidal categories allow us to transfer
additional structures given on objects of $\cA$ to objects from
$\wt{\cA}$. For example, let $X\in Ob(\cA)$.
A morphism will
be called $\cA$-associative (or $\phi$ associative)
 if we have the following equality of
morphisms of $(X\t X)\t X\to X$
$$\mu(\mu\t id)=\mu(id\t\mu)\phi.$$
Then, for $\al(X)\in Ob(\wt{\cA})$ the naturally
defined  morphism $\al(\mu)\beta^{-1}:\al(X)\t\al(X)\to\al(X)$
will be $\wt{\cA}$-associative ($\tilde\phi$-associative).

Let $A$ be a commutative algebra with unit, $B$ a unitary
$A$-algebra. The category of representations of $B$ in $A$-modules,
i.e. the category of $B$-modules, will be a monoidal category
if the algebra $B$ is equipped with additional structures \cite{Dr}.
Suppose we have  an algebra morphism, $\De: B\to B\t_A B$,
which is called a comultiplication,
and $\Phi\in B^{\t 3}$ is an invertible element such that $\De$ and $\Phi$
satisfy the conditions
$$(id\t\De)(\De(b))\cdot\Phi=\Phi\cdot (\De\t id)(\De(b)),\ \ b\in B, \tag{3}$$
$$(id^{\t 2}\t \De)(\Phi)\cdot(\De\t id^{\t 2})(\Phi)=
(1\t\Phi)\cdot(id\t\De\t id)(\Phi)\cdot(\Phi\t 1).         \tag{4}$$

We define a tensor product functor
which we will denote  $\t_{\cA}$ for $\cA$ the category of
$B$ modules or simply $\t$ when there can be no confusion in
the following way: given $B$-modules $M,N$
$M\t_{\cA} N=M\t_A N$ as an $A$-module with the
action of $B$ defined as $b(m\t n)=b_1m\t b_2n$ where $b_1\t b_2=\De(b)$.
The element $\Phi$ gives an associativity constraint
$\Phi:(M\t N)\t P\to M\t(N\t P),
(m\t n)\t p\mapsto \Phi_1m\t(\Phi_2n\t\Phi_3p)$, where
$\Phi_1\t\Phi_2\t\Phi_3=\Phi$.
By virtue of (3) $\Phi$ induces an isomorphism of $B$-modules, and by
virtue of
(4) the pentagonal identity (1) holds. We call the
triple
$(B,\De,\Phi)$ a Drinfeld algebra.
Thus, the category $\cA$ of $B$-modules for $B$ a
Drinfeld algebra  becomes a monoidal category. When
it becomes necessary to be more explicit  we shall denote
${\cA}(B,\De,\Phi)$.

Let $(B,\De,\Phi)$ be a Drinfeld algebra and $F\in B^{\t 2}$
an invertible element. Put
$$\wt{\De}(b)=F\De(b)F^{-1},\ \ b\in B,  \tag{5}$$
and
$$\wt{\Phi}=(1\t F)\cdot(id\t\De)(F)\cdot\Phi\cdot(\De\t id)(F^{-1})
\cdot(F\t 1)^{-1}.           \tag{6}$$
Then $\wt{\De}$ and $\wt{\Phi}$ satisfy (3) and (4),
therefore the triple $(B,\wt{\De},\wt{\Phi})$ also becomes a Drinfeld
algebra which generates the corresponding monoidal category
$\wt{\cA}(B,\wt{\De},\wt{\Phi})$. Note that the categories $\cA$
and $\wt{\cA}$ consist of the same objects as $B$-modules,
and the tensor products of two objects are isomorphic as $A$-modules.
The categories $\cA$ and $\wt{\cA}$ will be equivalent. The equivalence
$\cA\to \wt{\cA}$ is given by the pair $(\al,\beta)=(Id,F)$,
where $Id:\cA\to \wt{\cA}$ is the identity functor of the categories
(considered without the monoidal structures, but only as categories
of $B$-modules), and $F:M\t_{\cA} N\to M\t_{\wt{\cA}} N$ is defined by
$m\t n\mapsto F_1m\t F_2n$ where $F_1\t F_2=F$.
By virtue of (5) $F$ gives an isomorphism of $B$-modules, and (6)
implies the commutativity of diagram (2).

Assume $M$ is a $B$-module with a multiplication $\mu:M\t_A M\to M$
which is a homomorphism of $A$-modules. We say that $\mu$
is invariant with respect to $B$ and $\De$ if it is a morphism
in the monoidal category $\cA(B,\De,\Phi)$. This means that
$$b\mu(x\otimes y)=\mu\De(b)(x\t y)\ \ \ \mbox{for}\ b\in B,\ x,y\in M.
\tag{7}$$
When $\mu$ is $\cA$-associative, $\cA=\cA(B,\De, \Phi)$, then
we shall also say that $\mu$ is
a $\Phi$-associative multiplication, i.e. we have the equality
$$\mu(\mu\t id)(x\t y \t z)=\mu(id\t\mu)\Phi(x\t y \t z)\ \ \ \mbox{for}\
x,y,z\in M. \tag{8}$$
Since the pair $(Id,F)$ realizes an equivalence of the categories,
the multiplication $\wt{\mu}=\mu F^{-1}:M\t_A M\to M$ will be
$\wt{\Phi}$-associative and invariant in the category $\wt{\cA}$.

Now we return to the situation of Section \ref{s1}.
Let $\g$ be a semisimple Lie algebra over $\bC$ with a fixed Cartan
decomposition and an involution $\sigma$. Let $U\g$ be the universal
enveloping algebra with the usual comultiplication $\De:U\g\to U\g^{\t
2}$ generated  as a morphism of algebras by the equations
$\De(x)=1\t x+x\t 1$ for $x\in\g$ and extended multiplicatively.

We will deal with the category $\R(U\g[[h]])$. Objects of this category
are representations of $U\g\bh$ in $\bC\bh$-modules of the form
$E\bh$ for some vector space $E$. We denote here by $E\bh$ the set of
formal power series in an indeterminate $h$ with coefficients in
$E$. By tensor product of two $\bC[[h]]$-modules we mean the completed
tensor product in $h$-adic topology, i.e. for two vector spaces
$E_1$ and $E_2$ we have $E_1\bh\t E_2\bh=(E_1\t_{\bC} E_2)\bh$.
As usual, morphisms in this category are morphisms of $\bC\bh$-modules
that commute with the action of $U\g\bh$. A representation  of $U\g\bh$
on $E\bh$ can be given by a power series
$R_h=R_0+hR_1+\cdots+h^nR_n+\cdots\in End(E)\bh$ where
 $R_0$ is a $\bC$ representation of $U\g$ in $E$ and $R_i\in
\mbox{Hom}_{\bC}(U(\g), End(E))$. Hence, $R_h$ may be
considered as a deformation of $R_0$. By misuse of language, we will
say that $R_h$ is a representation of $U\g$ in the space $E[[h]]$.
The functor $\t_{\bC} \bC[[h]]$ sending a representation of $U\g$ to a
representation of $U\g[[h]]$ defines an equivalence of categories
between the category $\R(U\g)$ of representations of $U\g$ and the
category $\R(U\g[[h]])$ so we will shorten notation denote the latter by
$R(U\g)$ as well.

Since the comultiplication $\De$ on $U\g$ gives rise to a
comultiplication on $U\g\bh$ and is coassociative, the triple
$(U\g\bh,\De,1\t 1\t 1={\bf 1})$ becomes a Drinfeld algebra and the category
$\R(U\g)$ turns into a monoidal category $\R(U\g,\De,{\bf 1})$
with the identity
associativity constraint. This is the classical way to introduce
a monoidal structure in the category $\R(U\g)$. Another possibility
arises from the theory of quantum groups due to
Drinfeld. In the following proposition we suppose that the element
$\ff=[r,r]_{Sch}$ is invariant under the involution $\sigma$.

\begin{proposition}
\label{p2.1}
1. There is an invariant element $\Phi_h\in U\g\bh^{\t 3}$ of the form
$\Phi_h=1\t 1 \t 1+h^2\ff+\cdots$  satisfying the following
properties:

\ a) it depends on $h^2$, i.e. $\Phi_h=\Phi_{-h}$;

\ b) it satisfies the equations (3) and (4) with the usual $\De$;

\ c) $\Phi^{-1}_h=\Phi^{321}_h$, where
$\Phi^{321}=\Phi_3\t\Phi_2\t\Phi_1$ for $\Phi=\Phi_1\t\Phi_2\t\Phi_3$;

\ d) $\Phi_h$ is invariant under the Cartan involution $\theta$ and $\sigma$;

\ e) $\Phi_h\Phi_h^{s}=1$, where
$s$ is the antipode, i.e., an antiinvolution of $U\g$ such that
$s(x)=-x$ for $x\in \g$, and $\Phi_h^{s}=(s\t s\t s)(\Phi_h)$.

2. There is an element $F_h\in U\g\bh^{\t 2}$ of the form
$F_h=1\t 1+hr+\cdots$ satisfying the following properties:

\ a) it satisfies the equation (6) with the usual $\De$ and with
$\wt{\Phi}=1\t 1\t 1$;

\ b) it is invariant under the Cartan subalgebra $\c$;

\ c) $F_{-h}=F^{\theta}_h=F^{21}_h$;

\ d) $F_h(F^s_h)^{21}=1$
\end{proposition}
\Proof Existence and properties a)--c) for $\Phi$ are proven by
Drinfeld \cite{Dr}. From his proof which is purely cohomological
it is seen that $\Phi$ can be chosen invariant under all those automorphisms
under which the element $\ff$ is invariant. This proves 1 d). Similarly
1 e) can  be deduced from the cohomological construction by
restricting to a  suitable subcomplex, \cite{DS}.

Existence and the property a) for $F$ are also proven by Drinfeld
\cite{Dr}. In his proof he used the explicit existence of the
Drinfeld--Jimbo quantum group $U_h\g$. A purely cohomological
construction of $F$, not assuming the existence of the Drinfeld--Jimbo
quantum group, and establishing the properties listed in 2 b)--2 d)
is given in \cite{DS}. \QED

So, we obtain two nontrivial Drinfeld algebras: $(U\g,\De,\Phi)$
with the usual comultiplication and $\Phi$ from Proposition
\ref{p2.1}, and $(U\g,\wt{\De},id)$ where $\wt{\De}(x)=F_h\De(x)F_h^{-1}$
for $x\in U\g$. The corresponding monoidal categories $\R(U\g,\De,\Phi)$
and $\R(U\g,\wt{\De},{\bf 1})$ are isomorphic, the isomorphism being given
by the pair $(Id,F_h)$. Note that the bialgebra $(U\g\bh,\wt{\De})$
is coassociative one and is isomorphic to Drinfeld-Jimbo quantum group
$U_h\g$ by Drinfeld's uniqueness theorem, for proof see \cite{SS}.
 So that the category $\R(U\g,\wt{\De},{\bf 1})$ with the trivial
associativity constraint is called the category of representation of
quantum group. Note that if we ``forget'' the monoidal
structures all three categories are isomorphic to the category
$\R(U\g)$.\\

{\bf Remark.}
Corresponding to the category $\R=\R(U\g,\De,\Phi)$
define a category $\R'$ with the
reversed tensor product, $V\t' W=W\t V$, and the associativity
constraint $\Phi'((V\t' W)\t' U)=\Phi^{-1}(U\t(W\t V))$.
Denote by $S:V\t W\to W\t V$ the usual permutation, $v\t w\mapsto w\t v$,
which we will consider as a mapping $V\t W\to V\t'W$.
Then the condition 1 c) for $\Phi$ implies that the pair $(Id, S)$ defines
an equivalence of the categories $\R$ and $\R'$.

The antiinvolution $s$ defines an antipode
on the bialgebra $U\g$. The existence of the antipode
and property 1 e)
for $\Phi_h$ makes $\R$ into a rigid monoidal category.
The property 2 c) for $F_h$ gives an equivalence of the categories
$\R(U\g,\De,\Phi)$ and $\R(U\g,\wt{\De},{\bf 1})$ as rigid monoidal
categories (see \cite{DS} for more details).

\def\DM{{\rm Diff}(M)}
\def\op{\oplus}
\def\na{\nabla}

\section{Quantization}
\label{s3}
Let $A$ be the sheaf of smooth functions on a smooth manifold $M$.
Let $\DM$ be the sheaf of linear differential operators on $M$.
A $\bC$-linear mapping $\lambda:\t_{\bC}^n A\to A$ is called an
$n$-differential cochain if there exists an element $\hat{\lambda}\in
\t_A^n\DM$ such that
$\lambda(a_1\t \dots \t a_n)=\hat{\lambda}_1a_1\cdot\hat{\lambda}_2a_2
\cdots\hat{\lambda}_na_n$, where $\hat{\lambda}=\hat{\lambda}_1
\t\cdots\t\hat{\lambda}_n$ (summation understood).
It is easy to see that the element
$\hat{\lambda}$ is uniquely determined by the cochain $\lambda$.
We say that $\lambda$  is ``null on constants", if
$\lambda(a_1\t \dots \t a_n)=0$ in case at least one of $a_i$ is a constant.
Such $\lambda$ is presented by $\hat{\lambda}\in\t_A^n\DM_0$ where
$\DM_0$ denotes differential operators which are zero on constants.
 From now on we only consider $n$-differential cochains that are zero
on the constants. Denote by $H^n(A)$ the Hochschild cohomology
defined by the complex of such spaces.

It is known that the space $H^n(A)$ is isomorphic to the space
of the antisymmetric $n$-vector fields on $M$. Suppose that a group
$G$ acts on $M$ and there exists a $G$-invariant connection on $M$.
In this case Lichnerowicz proved (\cite{Li}) for $n\leq 3$ that
$H^n_G(A)$ is isomorphic to the space of the $G$-invariant antisymmetric
$n$-vector fields on $M$. Here $H_G(A)$ is the  cohomology of the
subcomplex of $G$-invariant cochains.

We will consider  cochains $\lambda_h:A\bh^{\t 2}\to A\bh$ given by
power series from $\DM^{\t 2}\bh$ of the form $\lambda_h=1\t 1+
\sum h^i\lambda_{1i}\t\lambda_{2i}$. ( By our convention, each $\lambda_{1i}\t
\lambda_{2i}$ is a sum over a second index $j$.) This means that
$\lambda_h(a\t b)=\sum_i h^i\lambda_{1i}(a)\lambda_{2i}(b),$
where $\lambda_0(a,b)=ab$.
We will also write $\lambda_h:A^{\t 2}\to A$. The cochain
$\mu_h:A^{\t 2}\to A$ is called equivalent to $\lambda_h$ if there
exists a differential $1$-cochain $\xi_h:A[[h]]\to A\bh,
\ \xi_h=1+\sum h^i\xi_i$ such that
$\mu_h(a\t b)=\xi_h^{-1}\lambda_h(\xi_h a\t \xi_h b)$,
where inverse is computed in the  sense of formal power series.

Let $M$ be a symmetric space, as in Section \ref{s1}. Consider
the space $A=\CM$ as an object of the category $\R(U\g,\De,\Phi_h)$
where $\Phi_h$ is from Proposition \ref{p2.1}.

\begin{proposition}
\label{p3.1}
There is a multiplication $\mu_h$ on $A$ with the properties:

a) $\mu_h$ is $\Phi_h$-associative, i.e.
$$\mu_h(\mu_h\t id)(a \t b \t c)=\mu_h(id\t\mu_h)\Phi_h(a \t b \t c),\ \
a,b,c\in A;$$

b) $\mu_h$ has the form
$$\mu_h(a\t b)=ab+\sum_{i\geq 4}h^i\mu_i(a \t b),$$
where $\mu_i$ are two-differential cochains, null on constants.
Moreover, $\mu_h=\mu_{-h}$, i.e., $\mu_h$ depends only on $h^2$.

c) $\mu_h$ is invariant under $\g$ and $\tau$;

d) $\mu_h$ is commutative, i.e.
$$\mu_h(a\t b)=\mu_h(b \t a).$$

The multiplication with such properties is unique up to equivalence.
\end{proposition}
\Proof We use arguments from \cite{Li}, proceeding by induction.
We may put $\mu_1=\mu_2=0$, because
the usual multiplication $m(a\t b)=ab$ satisfies a) modulo $h^4$.
This follows because $\Phi_h$ is a series in $h^2$ and the
$h^2$-term $\ff=0$ on $M$.
Suppose we have constructed $\mu_i$ for even $i<n$, such that
$\mu^n_h=\sum'\mu_i h^i$ satisfies a)--d) modulo $h^n$, where $\sum'$
denotes sum over even indices.
Then,
$$\mu_h^n(\mu_h^n\t id)=\mu_h^n(id\t\mu_h^n)\Phi_h+h^n\eta\
\mbox{mod} h^{n+2},
\tag{9}$$
where $\eta$ is an invariant three-cochain.

The following direct computation using the pentagon identity for $\Phi_h$ shows
that $\eta$ is a Hochschild cocycle.
By definition
$$d\eta= m(id\t \eta) -\eta (m\t id^{\t 2}) +\eta(id\t m\t id)
-\eta(id^{\t 2} \t m) +m(\eta \t id).$$
Using (9)  and calculating modulo $h^{n+2}$ we can
replace $m$ with $\mu^n_h$. Furthermore, the $G$-invariance of
$\mu^n_h$ implies that
\begin{eqnarray*}
\Phi(\mu^n_h\t id ^{\t 2})&=&(\mu^n_h\t id^{\t 2})(\Delta \t id^{\t 2})\Phi,\\
 \Phi(id\t \mu^n_h \t id)&=&(id\t \mu^n_h \t id)
( id\t \Delta\t id )\Phi, \\
\Phi( id^{\t 2}\t \mu^n_h)&=&( id^{\t 2}\t \mu^n_h)( id^{\t 2} \t \Delta )\Phi.
\end{eqnarray*}
Therefore we have the following equations modulo $h^{n+2}$,
\begin{eqnarray*}
\mu^n_h(id\t\mu_h^n)(id\t\mu_h^n\t id)-
\mu^n_h(id\t\mu_h^n)(id^{\t 2} \t\mu_h^n)(1\t\Phi_h)\quad\\
=h^n m(id\t\eta)\\
\mu_h^n(\mu_h^n\t id)(\mu^h_n\t id^{\t 2} )-\mu_h^n(id\t\mu_h^n)
(\mu^n_h\t id^{\t 2})(\Delta \t id^{\t 2})(\Phi_h)\quad\\
=h^n\eta(m\t id^{\t 2})\\
\mu_h^n(\mu_h^n\t id)(id\t \mu^n_h\t id)-
\mu_h^n(id\t\mu_h^n)(id\t \mu^n_h\t id)(id\t \Delta \t id)(\Phi_h)\quad\\
=h^n\eta(id\t m \t id)\\
\mu_h^n(\mu_h^n\t id)(id^{\t 2}\t \mu^n_h)-\mu_h^n(id\t\mu_h^n)
(id^{\t 2}\t \mu^n_h)(id ^{\t 2}\t \Delta)(\Phi_h)\quad \\
=h^n\eta(id ^{\t 2}\t m)\\
\mu^n_h(\mu_h^n\t id)(\mu_h^n\t id^{\t 2})-
\mu^n_h(\mu_h^n\t id)(id\t\mu_h^n\t id)(\Phi_h\t 1)\quad\\
=h^n m(\eta\t id).
\end{eqnarray*}

Since the equations are congruences modulo $h^{n+2}$ and
$h^n \Phi =h^n\quad\mbox{mod} h^{n+2}$ the equations remain
valid if we multiply on the left by any expression in $\Phi$ and
leave the right side unchanged.
 Multiply the left side of the first equation  by $((id\t \Delta \t id)\Phi)
(\Phi \t 1)$, the left side of the third equation by $\Phi \t 1$,
the left side of the fourth equation  by $(\Delta \t id \t id)\Phi$,
 leave the remaining equations unchanged, then add the five equations
with alternating signs.
Using the pentagon identity in $\Phi$ and the identity
$ (\mu^n_h\t id)(id \t id \t \mu^n_h)=\mu^n_h\t\mu^n_h=
(id \t \mu^n_h)(\mu^n_h \t id \t id),$ we conclude that $d\eta=0.$

Since $\g$ is semisimple the cochains
  invariant under $\g$ and $\dot\tau$ form a subcomplex which
is a direct summand.
The arguments from the proof
of Proposition \ref{p1.1} show that there are no three-vector fields on $M$
invariant under $\g$ and $\dot\tau$ .
Hence the cohomology of this subcomplex  is equal to zero, i.e. $\eta$
is a coboundary. Further, there is a $\g$ and $\tau$ invariant
connection on $M$ (see \cite{He} 4.A.1). The property
$\Phi_h^{-1}=\Phi_h^{321}$
 and commutativity of $\mu_i$ imply that
$\eta(a\t b\t c)=\eta(c\t b \t a)$. It follows  that there is
an invariant commutative two-cochain $\mu_n$ such that $d\mu_n=\eta$,
which shows that
$\mu^n_h+h^n\mu_n$ satisfies  a)--d) modulo $h^{n+2}$.
Therefore, proceeding step-by-step we can build the multiplication
$\mu_h$.

 The equivalenceof  any two such multiplications follows from the fact
that any symmetric Hochschild differential-two-cochain bounds.\QED

Now we suppose that on the algebra $A$ there is a $\g$ and $\tau$ invariant
multiplication $\mu_\nu:(A\t A)[[\nu]]\to A[[\nu]]$ which is associative
in the usual sense and such that $\mu_0=m$ where $m$ is the usual
multiplication on $A$.
The multiplication $\mu_\nu$ exists when $M$ is a k\"{a}hlerian symmetric
space. In this case $\mu_\nu$ can be constructed as the deformation
quantization of the Poisson bracket $\{\cdot,\cdot\}_{inv}$
which is the dual to the
k\"{a}hlerian form on $M$. Such a quantization also can be constructed
using the arguments of Proposition \ref{p3.1} and has the form
$$\mu_\nu(a,b)=ab+\frac{1}{2}\nu\br{a}{b}_{inv}+o(\nu).$$
Moreover, $\mu_\nu$ satisfies the property
$$\mu_\nu(a,b)=\mu_{-\nu}(b,a).$$
Denote by $A_\nu$ the corresponding algebra.
Let $H^n(A_\nu)$ be the Hochschild cohomology of this algebra.
Since $H^3_{G,\tau}(A_0)=0$ it is easy to see that
$H^3_{G,\tau}(A_\nu)=0$
as well. Using the same arguments as in the proof of
Proposition \ref{p3.1}, we have the following

\begin{proposition}
\label{p3.2}
Let $M$ be a k\"{a}hlerian symmetric space, $\mu_\nu$ the quantization
of the Kirillov bracket. Then
there is a multiplication $\mu_{\nu,h}$ on $A=\CM$ depending on
two formal variables with the properties:

a) $\mu_{\nu,h}$ is $\Phi_h$-associative, i.e.
$$\mu_{\nu,h}(\mu_{\nu,h}\t id)(a \t b \t c)=
\mu_{\nu,h}(id\t\mu_{\nu,h})
\Phi_h(a \t b \t c),\ \ a,b,c\in A;$$

b) $\mu_{\nu,h}$ has the form
$$\mu_{\nu,h}(a\t b)=\mu_{\nu}(a \t b)+\sum_{i\geq 4}h^i\mu_{\nu,i}(a\t b),$$
where $\mu_{\nu,i}:(A\t A)[[\nu]]\to A[[\nu]]$ are 2-differential cochains
 null  on constants.
Moreover, $\mu_{\nu, h}$ depends only of $h^2$,
i.e. $\mu_{\nu,h}=\mu_{\nu,-h}$, and
$\mu_{\nu,h}(a,b)=\mu_{-\nu,h}(b,a)$.

c) $\mu_{\nu,h}$ is invariant under $\g$ and $\tau$;

d) $\mu_{\nu,0}$ coincides with $\mu_\nu$, and
$\mu_{0,h}$ coincides with $\mu_h$ from Proposition \ref{p3.1}.

The multiplication with such properties is unique up to equivalence.
\end{proposition}

Now let us consider $A=\CM$ as an object of the category
$\R(U\g,\tilde{\De},{\bf 1})$ of representations of the Drinfeld--Jimbo
quantum group $U_h\g$. As we have seen in Section \ref{s2}, the
multiplications $\mu_h$ and $\mu_{\nu,h}$ can be transfered
to this category in the following way:
\def\ti{\tilde}
$$\ti{\mu}_h=\mu_hF^{-1}_h$$
$$\ti{\mu}_{\nu,h}=\mu_{\nu,h}F^{-1}_h.$$
We may obviously assume that $F_h$ has the form
$$F_h=1\t 1-\frac{1}{2}h\br{\cdot}{\cdot}+o(h).$$

Then we have the following
\begin{theorem}
\label{t3.3}
Let $M$ be a symmetric space over a semisimple Lie group.
Then the multiplications $\ti{\mu}_h$ and $\ti{\mu}_{\nu,h}$
(the second exists when $M$ is a k\"{a}hlerian symmetric space)
satisfy the following properties:

a) $\ti{\mu}_h$ and $\ti{\mu}_{\nu,h}$ are associative;

b) $\ti{\mu}_h$ and $\ti{\mu}_{\nu,h}$ have the form
$$\ti{\mu}_{h}(a\t b)=ab+\frac{1}{2}h\br{a}{b}+o(h)$$
$$\ti{\mu}_{\nu,h}(a \t b)=
ab+\frac{1}{2}(h\br{a}{b}+\nu\br{a}{b}_{inv})+o(h,\nu)$$

c) $\ti{\mu}_h$ and $\ti{\mu}_{\nu,h}$ are invariant under action
of the Drinfeld--Jimbo quantum group $U_h\g$;

d) $\ti{\mu}_{\nu,0}$ coincides with $\mu_\nu$, and
$\ti{\mu}_{0,h}$ coincides with $\ti{\mu}_h$;

e) Let $\ti{S}=F_hSF_h^{-1}$ where $S$ denotes the usual
transposition, $S(a\t b)=b\t a$ for $a,b\in A$.
Then  $\ti{\mu}_h$ is $\ti{S}$-commutative:
$$\ti{\mu}_h(a\t b)=\ti{\mu}_h\ti{S}(a \t b)\ \ \mbox{for } a,b \in \CM.$$
For $\ti{\mu}_{\nu,h}$ one has:
$$\ti{\mu}_{\nu,h}(a\t b)=\ti{\mu}_{-\nu,h}\ti{S}(a \t b)\ \
\mbox{for } a,b \in \CM.$$
The multiplications with such properties are unique up to equivalence.
\end{theorem}

{\bf Remarks.}

1. The action of the real Lie group $G$ and $\tau$ on $M$
induces an action on $\CM\bh$. It follows from
Propositions \ref{p3.1} and \ref{p3.2}
that $\mu_h$ and $\mu_{\nu,h}$ are invariant under $G$ and $\tau$.
This implies that $\ti{\mu}_{h}$ and $\ti{\mu}_{\nu,h}$ will be
invariant under a ``quantized'' action of $G$ and $\tau$.
This new action appears by taking of tensor products of $\CM$.
Namely, let $g$ be either an element of $G$ or $g=\tau$, then for
$a,b\in \CM$
define $g\circ_ha=g\circ a$, $g\circ_h(a\t b)=F_h(g\t g)F_h^{-1}(a\t
b)$, where $\circ$ denotes the usual action. The multiplications
$\ti{\mu}_{h}$ and $\ti{\mu}_{\nu,h}$ are invariant under this
quantized action, i.e., for example,
$$g\circ_h\ti{\mu}_{\nu,h}(a,b)=\ti{\mu}_{\nu,h}g\circ_h(a\t b).$$

2. We may consider a complex symmetric space $M=G/H$, where $G$ is
a complex semisimple Lie group and $H$ a complex subgroup. As above, one can
construct the multiplications $\mu_h$ and $\ti{\mu}_h$ on the space
$\CM$ that also will give a multiplication on the space of
holomorphic functions on $M$. The previous remark remains valid for
the complex group $G$.

In particular, the group $G$ itself may be considered as a symmetric
space, $G=(G\times G)/D$ where $D$ is the diagonal.
The action of $G\times G$ on $G$ is $(g_1,g_2)\circ g=g_1gg^{-1}_2$,
$(g_1,g_2)\in G\times G,\ g\in G$.
In this case
$\sigma(g_1,g_2)=(g_2,g_1)$, $\tau(g)=g^{-1}$.
In order for $\ff$ to be $\sigma$-invariant the corresponding
$R$-matrix can be taken in the form
$\bar{r}=(r,r)\in \wedge^2\g_1\oplus\wedge^2\g_2
\subset\wedge^2(\g_1\oplus\g_2)$ or $\ti{r}=(r,-r)$,
where the Lie subalgebras $\g_1$ and
$\g_2$ correspond
to $(G\times 1)$ and $(1\times G)$. In this example
$U\bar{\g}=(U\g)^{\t 2}\supset U\g\oplus U\g$ and in the both cases the
element $\bar{\Phi}_h$ has the form $\bar{\Phi}_h=(\Phi_h,\Phi_h)$
with $\Phi_h$ from Proposition 3.1.
In case $\bar{r}$ the corresponding
$\bar{F}_h$ has the form
$\bar{F}_h=(F_h,F_h)$ with $F_h$ from Proposition 3.1.
In case $\ti{r}$ the corresponding
$\ti{F}_h$ has the form
$\ti{F}_h=(F_h,F_{-h})$ with $F_h$ from Proposition 3.1.
Then, $\rho(\bar{\Phi}_h)=id$, so that for $\mu_h$ one can take the usual
multiplication $m$
on $C^\infty(G)$, and $\ti{\mu}_h(a,b)=m(F_h(a\t b)F_h^{-1})$ in the
case $\bar{r}$, and $\ti{\mu}_h(a,b)=m(F_h(a\t b)F_{-h}^{-1})$ in the
case $\ti{r}$.
Therefore, in the both cases $C^\infty(G)$ may be considered as an algebra
in the
category $\R((U\g)^{\t 2},\ti{\De},{\bf 1})$ with the multiplication
$\ti{\mu}_h$.
Note that the first multiplication is a quantization of the Poisson
bracket $(r-r')$ on $G$ where $r$ and $r'$ denote the extensions of $r$
as right- and left-invariant bivector fields on $G$, whereas
the second multiplication is a quantization of the Poisson bracket
$(r+r')$ on $G$.

\end{document}